Comment on « Locating the source field lines of Jovian decametric radio emissions » by YuMing Wang et al.

Laurent Lamy[1,2,3], Baptiste Cecconi[1,2], Stéphane Aicardi[4], C. K. Louis[5]

[1]LESIA, Observatoire de Paris, Université PSL, CNRS, Sorbonne Université, Université de Paris, 5 place Jules Janssen, 92195 Meudon, France.
[2]Station de Radioastronomie de Nançay, Observatoire de Paris, Université PSL, CNRS, Univ. Orléans, 18330 Nançay, France.
[3]LAM, Pythéas, Aix Marseille Université, CNRS, CNES, 38 Rue Frédéric Joliot Curie, 13013 Marseille, France.
[4]DIO, UMS2201 CNRS, Observatoire de Paris, Université PSL, 61 avenue de l'Observatoire, 75014, Paris, France.
[5]School of Cosmic Physics, DIAS Dunsink Observatory, Dublin Institute for Advanced Studies, Dublin, Ireland.

**Abstract**
In this comment of the article "Locating the source field lines of Jovian decametric radio emissions" by YuMing Wang et al., 2020, we discuss the assumptions used by the authors to compute the beaming angle of Jupiter's decametric emissions induced by the moon Io. Their method, relying on multi-point radio observations, was applied to a single event observed on 14[th] March 2014 by Wind and both STEREO A/B spacecraft from ~5 to ~16 MHz, and erroneously identified as a northern emission (Io-B type) instead of a southern one (Io-D type). We encourage the authors to update their results with the right hemisphere of origin and to test their method on a larger sample of Jupiter-Io emissions.

**Introduction**
In a study published in 2020, (Locating the source field lines of Jovian decametric radio emissions, YuMing Wang, XianZhe Jia, ChuanBing Wang, Shui Wang, and Vratislav Krupar, 2020, hereafter W20) proposed a method based on multi-point radio observations of Jovian decametric emissions induced by Io (hereafter Io-DAM) to accurately locate the position of Io-DAM radiosources, their host magnetic field line and the wave emission angle at the source. Adopting the formalism proposed by Hess et al. (2008), replicated in Equation 2 of W20, the authors then converted the wave emission angle into the kinetic energy of electrons driving the radiation through the Cyclotron Maser Instability, assuming a loss cone electron distribution as its free energy source. The proposed method is interesting with broad implications. With respect to past studies, the authors newly used the up-to-date JRM09 magnetic field model computed from Juno in situ data (Connerney et al., 2018), complemented by a current sheet model still based on Voyager in situ data (Connerney et al., 1981), to minimize uncertainties in the calculated quantities.

To illustrate/validate their method, W20 applied it in their section 3 to a single case of Io-DAM emission, observed nearly simultaneously on 14[th] March 2014 by the radio instruments onboard three space probes, namely Wind/Waves (Bougeret et al., 1995) and STEREO-A and -B/Waves (Bougeret et al., 2008), from ~5 to ~16 MHz. These observations are shown in Fig. 1, replicated from Fig. 1 of W20. Assuming that the Io-DAM source region lies in the north (hence corresponding to the Io-B class of emission), W20 determined that the footprint of the magnetic field line hosting the radio emission lies ahead of the average Io UV footprint (Bonfond et al., 2009) by a 32° lead angle. They derived roughly constant emission angles within 61.5-63.5°, from which they obtained electron kinetic energies varying within 12.5-18 keV. Comparing these values to those published by Zarka (1998) and Hess et al. (2008), the authors concluded that their method is thus "*valid and reliable*".

We nonetheless disagree with the main assumption of W20 on the hemisphere of origin, which directly affects the results, and comment other assumptions used by the authors, as discussed below.

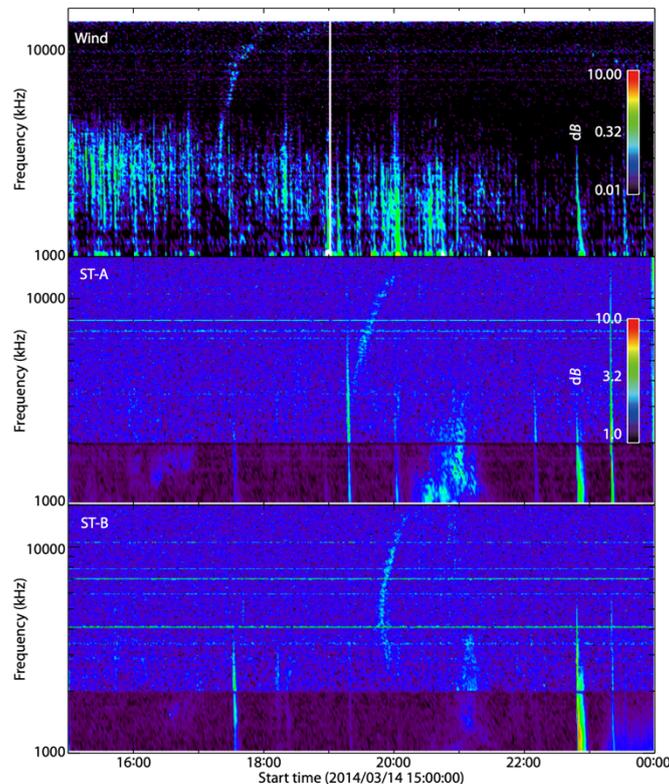

Fig. 1 : Io-DAM emission successively observed by Wind/Waves, and by STEREO-A and -B/Waves from ~5 to ~16 MHz (replicated from Fig. 1 of W20).

**Hemisphere of origin for the Io-DAM emission observed 14th March 2014**
To proceed to the analysis of the Io-DAM arc observed on 14th March 2014, the authors claim that the emission originates from the northern hemisphere on the basis of the following statement : "*We do not have the polarization measurements of the event, but on 2014 March 14 the northern magnetic pole was tilted toward the spacecraft. Thus, we believe that the DAM emission should come from the northern hemisphere.*" However, several arguments can be opposed to this identification and support that the investigated Io-DAM emission is not an Io-B arc (northern westward source) but an Io-D one (southern westward source).

It can be first seen that the radio arc observed by Wind/Waves displays an inflexion point near 10 MHz around 18:30 UT which is characteristic of Io-D arcs. Illustrations of the typical shape of Io-DAM arcs, observed over their full frequency range by Wind/Waves (<13 MHz) and the Nançay Decameter Array (NDA, >10MHz), can be found in (Queinnec and Zarka, 1998) (see their Fig. 1 for the Io-D case).

The Jovian magnetic latitude of the Earth, applicable to the Wind spacecraft, was varying between -8.3° and -4.1° from 17:00 UT to 19:00 UT, so that southern Io-DAM sources were in more favorable view from Wind/Waves. Similarly, the arcs observed by STEREO-A and -B/Waves between 18:20 and 19:20 UT corresponded to southern magnetic latitudes varying between -9° and -6°. In a recent paper investigating the statistical distribution of Jovian DAM emissions observed by Juno/Waves as a function of magnetic latitude, Louis et al. (2021) showed that the northern Io-B/A arcs, reaching higher frequencies than the Io-D/C ones owing to the larger northern magnetic field amplitude, are not observed below -5° magnetic latitude (see their Fig. 3d).

The classification of Io-DAM arcs in categories can be efficiently counter-checked from the visualization of the interval of observation on top of a classical CML-Io phase occurrence diagram, such as those built by Marques et al. (2017), Zarka et al. (2018) and refs therein, from a catalogue of 26 years of NDA observations. As illustrated on Fig. 2, on 14th March 2014, between 17:00 and 19:00 UT, Earth-based Wind/Waves observations intercepted the maximal probability region for Io-D emissions, only.

While STEREO/Waves is capable of polarization measurements (Cecconi et al., 2008), polarization measured during this event could only be tracked up to ~2 MHz, below which the Io-DAM arc was not visible (as opposed to the hectometric burst observed near ~21:00 UT). For the sake of completeness, we alternately checked the NDA public observations of Jupiter over 10-40 MHz obtained simultaneously to the Wind/Waves ones below 13 MHz on 14th March 2014. We could not identify any clear Jovian signature with either Right-handed or (expected) Left-handed polarization in the 10-16 MHz NDA range, where the band was unfortunately both strongly contaminated by RFIs (up to 18 MHz) and subject to low frequency filtering (below 14 MHz). Nonetheless, any intense enough Io-B emission should have reached frequencies >20 MHz (Queinnec & Zarka, 1998, Marques et al., 2017).

Finally, the lead angle of 32° obtained from the longitudinal difference between the footprint of the retrieved flux tube and that of the average Io UV spot was very large and, assuming that the Io-DAM sources lie along the main Io UV footprint, was not consistent with the results of Bonfond et al. (2009).

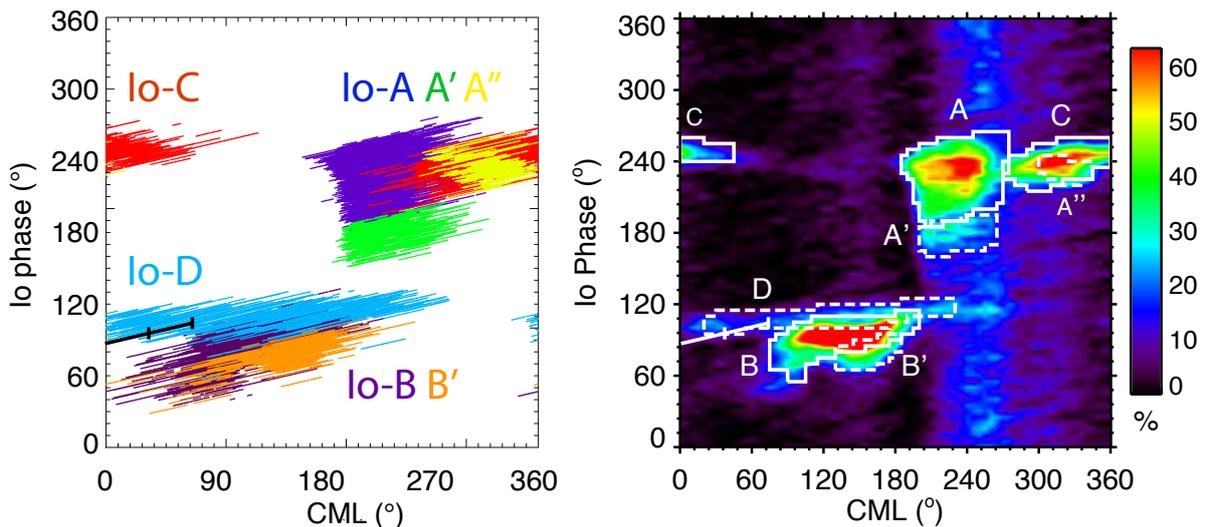

Fig. 2 : Plot of the Central Meridian Longitude (CML) as a function of Io phase, showing (a) the occurrence of Io-DAM arcs and (b) the occurrence probability of all Jovian DAM emissions derived from 26 years of NDA observations of Jupiter (Marques et al., 2017, Zarka et al., 2018). The solid line indicates the path of Earth-based observations on 14th March 2014 from 17:00 to 19:00 UT, when Wind/Waves recorded the Io-DAM arc displayed in Fig. 1. The line precisely intercepts the Io-D region, only. Both panels were created with the online

Jupiter probability tool, developed by the NDA team, at https://jupiter-probability-tool.obspm.fr.

**Other physical hypotheses**
To use the formalism of Hess et al. (2008), the authors define $f_{ce,max}$ (the maximal electron cyclotron frequency reached along the flux tube) as follows : "*The value of $f_{ce,max}$ is set to be the value at the footprint of each field line on the 1/15.4 flattened surface of one RJ, as an approximation of the frequency at the top of ionosphere.*" We note that the peak emission altitude of ~900 km for the Io UV footprint (Bonfond et al., 2009) would yield a more realistic value of $f_{ce,max}$. The parametric study of Io-DAM simulations with the ExPRES code of Louis et al. (2017) has quantified the effect of the altitude on the final time-frequency shape of the arc (see their Fig. 1e).

The relevance of the formalism of Hess et al. (2008) was not at all discussed by W20, while recent Juno in situ measurements of real electron distribution functions driving Jovian radio emissions provided the opportunity to assess its relevance (see e.g. Louarn et al., 2017, Louis et al., 2020).

Finally, the correspondence between the values plotted in left- and right-handed panels (which display different ranges along the y-axis) of Fig. 8 of W20 was not obvious, especially for the spread of dots.

**Conclusion**
We kindly encourage the authors to take into account the above comments to update their analysis, quantify the uncertainty on the parameters derived with a wrong hemisphere of origin, and ideally extend their method to a larger set of Io-DAM emissions from both hemispheres in order to achieve more "*valid and reliable*" results.

**Acknowledgements**
We acknowledge the JPL Horizons and IMCCE/MIRIADE ephemeris services, together with Vratislav Krupar for providing STEREO/Waves high level polarization data. The Jupiter probability tool at https://jupiter-probability-tool.obspm.fr was developed by the NDA team in the frame of the MASER service supported by Paris Astronomical Data Centre (PADC) at Observatoire de Paris. The (unshown) NDA data mentioned in the article are publicly available at https://www.obs-nancay.fr/reseau-decametrique/. We thank the CNES and CNRS/INSU programs of planetology and heliophysics.